# HYDRODYNAMIC VACUUM SOURCES OF DARK MATTER SELF-GENERATION IN ACCELERATED UNIVERSE WITHOUT BIG BANG.

S. G. Chefranov*, E. A. Novikov**.

*Obuchov Institute for Atmospheric Physics of Russian Academy of Sciences, Moscow, Russia; e-mail: schefranov@mail.ru
**Institute for Nonlinear Science, University of California, San Diego, La Jolla, USA; e-mail: enovikov@ucsd.edu

**Abstract -** We have obtained a generalization of the hydrodynamic theory of vacuum in the context of general relativity. While retaining the Lagrangian character of general relativity, the new theory provides a natural alternative to the view that the singularity is inevitable in general relativity and in the theory of a hot Universe. We show that the macroscopic source-sink motion as a whole of the ordinary (dark) matter during production-absorption of particles by vacuum generates polarization ( determining the variability of the cosmological term in general relativity). We have removed the well-known problems of the cosmological constant by refining the physical nature of dark energy associated precisely with this hydrodynamically initiated variability of the vacuum energy density. A new exact solution of the modified general relativity equations that contains no free (fitting) parameter additional to those available in general relativity has been obtained. It corresponds to continuous and metric-affecting production of ultralight dark matter particles ( with mass $m_o = (\hbar/c^2)\sqrt{12\rho_0 k} \approx 3 \cdot 10^{-66} g$, $k$ is the gravitational constant) out of vacuum, with its density $\rho_0$ being retained constant during the exponential expansion of spatially flat Universe. This solution is shown to be stable in the regime of cosmological expansion in the time interval $-\infty < t < t_{\max}$, where $t = 0$ corresponds to present epoch and $t_{\max} = 2/3H_0 c\Omega_{0m} \approx 38 \cdot 10^9$ years with $\Omega_{0m}=\rho_0/\rho_c \approx 0.28$ ($H_0$ is the Hubble constant, $\rho_c$ is the critical density). For $t > t_{\max}$, the solution becomes exponentially unstable and characterizes the inverse process of dark matter particles absorption by the vacuum in the regime of contraction of the Universe. We consider admissibility of the fact that scalar massive photon pairs can be these dark matter particles. Good quantitative agreement of the indicated exact solution with cosmological observations (SnIa, SDSS-BAO and recently found reduction of acceleration of the expanding Universe) has been obtained.

## 1. Introduction

The problem of explaining the causes of an accelerated expansion of the Universe, the conclusion about which follows from various present-day cosmological observations, has existed for more then 10 years [1-7]. These data largely confirm the predictions of the hydrodynamic theory of vacuum in general relativity [8,9] regarding the macroscopic gravitational manifestation of the physical vacuum - the ground state of quantized fields (see [4,5]). Indeed, the standard $\Lambda CDM$ model [2-4] corresponds to comparison between the constant vacuum energy density $\varepsilon_d = \varepsilon_{d0} = const$ and the cosmological constant $\lambda_0 = 8\pi k \varepsilon_{d0}/c^4$ ($k$ is Cavendish's universal gravitational constant, $c$ is the speed of light in vacuum), which is considered in Gliner's theory [8]. However using this identification of the quantities $\varepsilon_d$ and $\lambda_0$ inevitably leads to the well-known problems of the cosmological constant [1,4,5] (see also [10-12]).

At the same time, according to [8], the correspondence between $\varepsilon_d$ and $\lambda_0$ is possible only under the condition $\varepsilon_d = \varepsilon_{d0} = const$, when the metric of the spacetime manifold is determined exclusively by vacuum-phase matter and not by ordinary[1]) matter different from the vacuum. It was shown [9] that the presence of an arbitrary small amount of ordinary (metric-affecting) matter leads to instability of the initial (with $\varepsilon_d = \varepsilon_{d0} = const$) state of the vacuum against the production of ordinary matter out of it with the transition to a state of expansion. The as yet open question about the determination of $\varepsilon_d$, whose variability makes possible the production of ordinary matter out of the vacuum, is also posed there (see Eq. (1) in [9], where the designation $-\mu$ rather than $\varepsilon_d$ is used).

However, only the formal models of dark energy and dark matter are known today, for which, in contrast to [8,9], neither the nature of these hypothetical substances nor the physical mechanism of their possible interaction is clear (see the reviews [2-4] and, e.g., [13,14] and references therein).

At the same time, the importance of developing the theory [8,9] (on the macroscopic gravitational properties of the physical vacuum in the context of general relativity) is related to the possibility of removing the mentioned above and other problems of cosmology and theoretical physics. For example, this hydrodynamic theory of vacuum in general relativity provides a natural alternative [9] to the well-known view [15] that the singularity is inevitable in general relativity and in the corresponding theories [16,17].

Here, we develop a new hydrodynamic approach to determining the specific form of the variable vacuum energy density $\varepsilon_d$. This approach retains the Lagrangian character of general relativity and generalizes the theory of [8,9]. Namely, the hydrodynamic theory of distributed sources-sinks [18,19] is used for a macroscopic description of the production of ordinary (dark) matter out of vacuum. This theory allows the variability of the cosmological term to be modeled using the stretching $\sigma$ (the covariant divergence of the 4-velocity for ordinary matter). $\sigma$ is the Lagrangian invariant for the dynamics of distributed sources-sinks [18,19]. In this paper, we suggest a modification of the general relativity equations that is obtained by introducing an additional term $\gamma\sigma^2$ ($\gamma$ is dimensionless constant) into the Lagrangian density of the action functional of general relativity. Using $\sigma^2$ seems natural for the variational derivation of the form of $\varepsilon_d$ whose variability, according to [9], should characterize the intensity of the macroscopic flow of dark matter being produced out of the vacuum. Note that a more general modification of the equations, presented below in formula (3'), was derived from somewhat different consideration without variational principle [19].

Introducing term $\gamma\sigma^2$ can be considered as a generalization of Sakharov's gravitational theory of vacuum quantum fluctuations [11], where the expansion of the Lagrangian scalar density in powers the scalar curvature $R$ is used. The term $\gamma\sigma^2$, as well as the third term in (3'), have the same dimension as that of $R$ and the same order of magnitude [19]. This makes it possible to describe the production of massive particles out of the vacuum on macroscopic scales. The contribution from the quadratic (in $R$ and Ricci tensor) terms considered in [11] (and in various versions of the quantum gravity theory [2,20-22]) is relatively small compared to the contribution from $\sigma^2$ for such scales.

Thus, following [19], we suggest taking into account the new metric-affecting hydrodynamic factor that is capable of governing the macroscopic effects of vacuum polarization through the source-sink motion as a whole of the ordinary (dark) matter. This motion should inevitably emerge during the metric-affecting production (absorption) of massive particles of this matter out of the vacuum as the Universe expands (contracts). In this case, the hydrodynamic polarization factor under consideration gives an effect dependent on the manifold as a whole, just as the corresponding polarization contribution from the manifold topology [23].

Note that the additional term $\gamma\sigma^2$ considered in the action functional cannot be reduced to any function $f(R)$. Therefore, here we establish a refinement of the well known theorem on the equivalence of quantum gravity $f(R)$ models and scalar field models [24, 25]. Below, in Section 5, we obtain a condition for the equivalence of the new exact solution at $\gamma = 1/3$ (see Eq. (42) in Section 4) and the new exact solution for a modification of the scalar field model [26]. In contrast to [26], we take into account a nonzero and, what is important, negative cosmological constant, $\lambda_0 < 0$. Note that in the model [26] and in the equivalent quantum semiclassical Parker–Fulling model [27], it is possible to describe the reverse effect of the production of particles out of the vacuum on the metric, just as is done for the hydrodynamic theory under consideration at $\gamma \neq 0$. Here, we determine the relationship between the mass $m_0$ of the ultralight dark matter particles being produced out of the vacuum and the constant (under cosmological expansion) density $\rho_0$ of this matter by comparing these exact solutions.

The hydrodynamic approach suggested in [19] and here allows not only global cosmological phenomena but also more local processes to be modeled. This distinguishes it from the well known [28] formal models of dark energy $\varepsilon_d$ applicable only for global cosmological scales. In the nonrelativistic limit (Section 3), no problematic corrections to Newton's law of gravitation (as, for example, in [29, 30]) arise in this case for the modification of the general relativity equations under consideration. In contrast to [29, 31], below we consider not just the gravitational self-action of the moving matter but also the possibility of its self-generation out of the vacuum through the source–sink macroscopic vacuum polarization (for more detail, see the next section). We show (Section 6) that the exact solution obtained at $\gamma = 1/3$ quantitatively agrees with model-independent observational data at low redshifts $z$.

## 2. Hydrodynamic gravitational vacuum polarization and a new modification of the general relativity equations

Consider the effect of the source–sink macroscopic motions of ordinary (dark) matter on the metric (paragraph 1), the possibility of vacuum polarization from more general types of motion (paragraph 2), and the approaches realized in this connection in [29,31], etc. (paragraph 3).

**1.** Let us replace the cosmological constant $\lambda_0$ with the function $\lambda = \lambda_0 - \gamma\sigma^2$ in the action functional of general relativity [32], where

$$\sigma = \frac{\partial u^i}{\partial x^i} + \frac{u^k}{2g}\frac{\partial g}{\partial x^k}$$

is the covariant divergence of the 4 velocity $u^i$ for an ideal fluid, $x^i = (\tau, x^\alpha)$, $\alpha = 1, 2, 3$, $\tau = ct$, $t$ is the time, and $g$ is the determinant of the metric tensor $g_{ik}$. To derive the gravitational field equations, consider the following extremum condition for the sum of the action functionals of the gravitational field $S_g$ and matter $S_m$ [25, 32]:

$$\delta(S_g + S_m) = -\frac{c^3}{16\pi k}\delta\int d^4x\sqrt{-g}\,(R + 2\lambda) + \frac{1}{2c}\delta\int d^4x\sqrt{-g}\,L_m = 0 \qquad 1$$

Here $R$ is the scalar curvature or the Ricci scalar and $L_m$ is the Lagrangian of an ideal fluid that corresponds to the energy–momentum tensor of an ideal fluid (dark matter) in the form $T_i^k = (p + \varepsilon)u_i u^k - \delta_i^k p$, where $p$ and $\varepsilon$ are its pressure and energy density, respectively. In this case, we have the following representation for $L_m$[25, 33]:

$$L_m = p(1 + g^{ik}u_i u_k) - \varepsilon(1 - g^{ik}u_i u_k).$$

The normalization condition $g^{ik}u_i u_k = 1$ should be taken into account only after the variation in (1) independently in all components of the metric tensor and the vector $u^i$ is not varied but is considered fixed [25]. Particularly,

$$\delta(\sigma^2) = 2\sigma u^k \delta(\frac{1}{2g}\frac{\partial g}{\partial x^k}).$$

In addition, we take into account the fact that [32]

$$\delta\frac{\partial g}{\partial x^k} = \frac{\partial}{\partial x^k}(\delta g), \quad \delta g = -g g_{ik}\delta g^{ik}.$$

As a result, we obtain the following modification of the general relativity equations from Eq. (1) at any $\gamma$:

$$R_i^k - \frac{1}{2}\delta_i^k R = \frac{8\pi k}{c^4}(+\delta_i^k \varepsilon_d) \equiv \frac{8\pi k}{c^4}\tilde{T}_i^k, \qquad 2$$

$$\varepsilon_d = \frac{c^4 \lambda_1}{8\pi k}, \quad \lambda_1 = \lambda_0 + \gamma(2u^k \frac{\partial \sigma}{\partial x^k} + \sigma^2). \qquad 3$$

At $\gamma = 0$, Eqs. (2) and (3) coincide exactly with the general relativity equations at a nonzero cosmological constant, $\lambda_0 \neq 0$, and correspond to the $\Lambda CDM$ model. Equations (2) and (3) coincide with the special case (at $\beta = 2\gamma$) of the equations from [19] in which the following representation is used instead of $\lambda_1$ in (2) and (3):

$$\lambda_N = \lambda_0 + \beta u^k \frac{\partial \sigma}{\partial x_k} + \gamma \sigma^2, \qquad 3'$$

where $\beta$ and $\gamma$ are dimensionless constants.

From (2) we have (see [32]):

$$\tilde{T}_{i;k}^k = 0, \quad T_{i;k}^k = -\frac{\partial \varepsilon_d}{\partial x_i} \qquad 4$$

The second equation from (4) determines the possibility of energy and momentum exchange between the vacuum and nonvacuum phases of the states of matter [9]. Indeed, $\tilde{T}_i^k$ in Eq. (2) is obtained when $\varepsilon + \varepsilon_d$ and $p - \varepsilon_d$ are substituted for $\varepsilon$ and $p$, respectively, in $T_i^k$. This corresponds to the equation of state $p = -\varepsilon_d$ typical of the vacuum-like phase of matter [8]. In a comoving frame of references (where $u^0 = 1$, $u^\alpha = 0$, $\alpha = 1,2,3$) it follows from (3) that

$$\varepsilon_d = \frac{c^4}{8\pi k}\{\lambda_0 - \gamma[\frac{3}{4}(\frac{\dot{g}}{g})^2 - \frac{\ddot{g}}{g}]\}. \qquad 5$$

The dot here and below denotes a derivative with respect to $\tau$. From (4) we obtain the following modification of the general relativity continuity equation:

$$\dot{\varepsilon} + \frac{\dot{g}}{2g}(\varepsilon + p) = -\dot{\varepsilon}_d, \qquad 6$$

where $\dot{\varepsilon}_d \neq 0$ at $\gamma \neq 0$ for $\varepsilon_d$ from (5).

Thus, ordinary (dark) matter can be produced out of vacuum whose gravitational polarization, according to (3) or (3'), is generated by the new source-sink hydrodynamic factor. In this case, $\varepsilon_d$ is based on clear physical views that follow from the hydrodynamic theory of distributed sources-sinks in the context of Lagrangian modeling - from the Lagrangian invariants to dynamics [18,19]. Therefore, in contrast to the existing models for the interaction between dark energy $\varepsilon_d$ and dark matter $\varepsilon$ (see, e.g.,[13,14]), there is no need for the additional kinematic equation that formally closes the system of equations for two unknown function $\varepsilon$ and $\varepsilon_d$. Our approach can be consider as a generalization of not only [8,9] but also [11], because the sources-sink effect ensures the presence of ordinary matter, disregarded in [11], and its hydrodynamic interaction with the vacuum state of matter on macroscopic scales.

**2.** The consequences from the specific and simple representation $\lambda = \lambda_0 - \gamma\sigma^2$ used in (1) will be considered in more details in the next sections. Here, we will point out the possibility of generalizing this representation of $\lambda$ term in (1) not only through the replacement of $\sigma^2$ in (1) with an arbitrary function of $\sigma^2$ but also by taking into account other types of motion of matter. Namely, we put:

$$\lambda = \lambda_0 - \Phi(q_1, q_2, q_3), \qquad 7$$

$$q_1 = u^i_{;j} u^j_{;i} = \sigma^2, \quad q_2 = u^i_{;j} u^j_{;i}, \quad q_3 = g^{ij} g_{lm} u^l_{;i} u^m_{;j}, \qquad 8$$

where $q_a (a = 1, 2, 3)$ are quadratic scalars constructed from the covariant derivatives [32] of velocity field $u^i$, while generic function $\Phi$ (of three arguments) has dimension $cm^{-2}$, like $\sigma^2$. With $\Phi = \gamma q_1 = \gamma\sigma^2$ we return to the previous simple representation. An additional introduction of the important topological helicity invariant [34] into (1) and (7) is also possible.

**3.** In [29], the representation of $\lambda$ in (1) was used in the form of a weighted sum of expressions like (8) but with a hypothetical neo-ether vector field $A^i$ instead of velocity field $u^i$. Since, in contrast with $u^i$, the field $A^i$ in [29] is not fixed, additional equations are derived for it in [29] by varying the action functional in components of this vector. The density of the Lagrangian for ordinary matter is not considered in [29], while the corresponding energy-momentum tensor of ordinary matter is not derived from the variational principle (as in this paper) but is formally introduced into the resultant equations. As has been noted above (in Section 1), the obtained in [29] equations are not reduced to Newton's law of gravitation in the nonrelativistic limit, in contrast to this paper (see Section 3).

In recent paper [31] the $\lambda$ term is considered as a generic function of the argument

$$\psi^2 = \psi_{ik}\psi^{ik}, \quad \psi_{ik} \equiv \nabla_i u_k. \qquad 9$$

Here $\nabla_i u_k$ is the covariant derivative of the vector $u^k$, which formally has the same meaning as that in this paper and, thus, $\psi^2$ formally equivalent to scalar $q_2$ in (8). However, the procedure of varying the corresponding action functional in [31], in contrast to this paper, includes the variability of the field $u^i$ when varying in metric tensor components. That causes the field $u^i$ in [31] to be actually equivalent to the field $A^i$ in [29] with all of problem that follow from this. In particular, the violation of local Lorentz invariance is pointed out in [29], because $A^i \neq 0$. Indeed, the relation between variation $\delta u^i$ and $\delta g^{ik}$ in the form $\delta u^i = -(\delta g^{ik}/2)u_k$ is derived in [31] based on the normalization condition $g^{ik}u_i u_k = 1$. If such a finite variation of vector $u^i$ is taken into account when varying in (1), then this leads not to the form of the energy-momentum tensor for the ideal fluid that describes ideal fluid noted above but to $T^i_k = -p\delta^i_k$. This representation for $T^i_k$, according to [8], is characteristic of the vacuum phase of matter with the equation of state $p = -\varepsilon_d$ rather than ordinary matter. As a result, the entire matter considered in [31] is in a vacuum-like state, for which, according to [8], the energy density is constant, $\varepsilon_d = \varepsilon_{d0} = const$, any frame of reference should be a comoving one and this is incompatible with the condition $u^\alpha \neq 0$.

Therefore, the surprising (for the authors of [31]) fact that the right-hand side of Eq. (4) and (6) becomes zero gets a natural interpretation due to actually complete absence of ordinary matter in the analysis of [31]. This is disregarded in [31], because the energy-momentum tensor in both [29] and [31] is not derived from the variational principle, as is done above by considering $L_m$ in (1), but is only introduced formally.

Note also [35], whose author suggests a modification of the energy-momentum tensor for a nonideal fluid with the presence of the first and second viscosities. This leads in [35] to the possibility of a nonsingular evolution of the Universe, which is also obtained below (see Section 4) for an ideal fluid with distributed sources-sinks in it.

**3. The weak-field limit, gravitational waves, and Newton's law of gravitation.**

In that section, we show that the generation and emission of week gravitational waves is possible when the vacuum and nonvacuum phases of the state of matter interact and when the new modification of the general relativity equations (2) and (3) is reduced to Newton's law of gravitation in the nonrelativistic limit.

**1.** Consider the following representation of the metric tensor for the weak-field limit [32]:

$$g_{ik} = g^{(0)}_{ik} + h_{ik} + O(h^2), \quad g = g^{(0)}[1 + h + O(h^2)], \quad h \equiv h^i_i, \quad h^0_0 = \frac{2}{c^2}\varphi, \quad h^\beta_\alpha = -\frac{2}{c^2}\varphi_1, \qquad 10$$

where $g^{(0)}_{\alpha\beta} = -\delta_{\alpha\beta}$, $g^{(0)}_{00} = 1$, and $g^{(0)}_{0\alpha} = 0$.

In contrast to (106.3) from [32], the possibility to generally assume that $\varphi_1 \neq \varphi$ is considered in Eq. (10), i.e., there are two independent potentials, $\varphi_1$ and $\varphi$, rather than one. In the limit $h_{ik} \ll g^{(0)}_{ik}$, the corresponding expression for the interval is

$$ds^2 = (1 + \frac{2\varphi}{c^2})c^2 dt^2 - (1 - \frac{2\varphi_1}{c^2})(dx^2 + dy^2 + dz^2).$$

From Eqs. (2) and (3), using the well-known expression (see (107.6) from [32])

$$R_i^k = \frac{1}{2}(\Delta - \frac{1}{c^2}\frac{\partial^2}{\partial t^2})h_i^k,$$

where $\Delta$ is the three-dimensional Laplace operator, we obtain:

$$(\Delta - \frac{1}{c^2}\frac{\partial^2}{\partial t^2})h_i^k = \frac{16\pi k}{c^4}(T_i^k - \frac{1}{2}T\delta_i^k) - \lambda_1\delta_i^k, \qquad 11$$

where $T = T_i^i = \varepsilon - 3p$. According to (10), we have

$$h = \frac{2}{c^2}(\varphi - 3\varphi_1), \; g = -(1+h), \; \sigma = \frac{\dot{g}}{2g} \approx \frac{\dot{\varphi} - 3\dot{\varphi}_1}{c^2}.$$

Then, from (11) it follows:

$$\Delta\varphi - \frac{1-2\gamma}{c^2}\frac{\partial^2\varphi}{\partial t^2} = \frac{4\pi k(\varepsilon + 3p)}{c^2} - \lambda_0 c^2 + 6\gamma\frac{\partial^2\varphi_1}{\partial t^2}, \qquad 12$$

$$\Delta(\varphi + \varphi_1) - \frac{1}{c^2}\frac{\partial^2(\varphi + \varphi_1)}{\partial t^2} = \frac{8\pi k(\varepsilon + p)}{c^2}. \qquad 13$$

This system describes the gravitational waves with two potentials, $\varphi$ and $\varphi_1$. It is obtained in the limit of low velocities when we may set $u^0 = 1$, $u^\alpha = 0$, as with derivation of Eq. (6) from (4). In this case, in addition to (12) and (13), from Eq. (6) we get (after allowance for (10) and only the linear (in $\varphi$ and $\varphi_1$) terms in (6))

$$\frac{\partial\varepsilon}{\partial t} = -\frac{\gamma}{4\pi k}\frac{\partial^3(\varphi - 3\varphi_1)}{\partial t^3}. \qquad 14$$

The importance of considering Eq. (14) is determined by the necessity of controlling the fulfillment of (6) in this approximation for Eqs. (2) and (3).

**2.** We use the equation of state $p = x\varepsilon$ to close the system (12)-(14). The solution of this system can be represented as

$$\varphi = \varphi_s(\mathbf{r}) + \tilde{\varphi}(\mathbf{r},t), \; \varphi_1 = \varphi_{1s}(\mathbf{r}) + \tilde{\varphi}_1(\mathbf{r},t), \; \tilde{\varphi},\tilde{\varphi}_1 \propto \exp(-i\mathbf{n}\mathbf{r} + i\omega t).$$

From (12)-(14) we then obtain

$$\varepsilon = \varepsilon_0(\mathbf{r}) - \frac{\gamma}{4\pi k}\frac{\partial^2(\varphi - 3\varphi_1)}{\partial t^2}, \qquad 15$$

$$\Delta\varphi_s = \frac{4\pi k}{c^2}(1 + 3x)\varepsilon_0 - \lambda_0 c^2, \qquad 16$$

$$\Delta(\varphi_s + \varphi_{1s}) = \frac{18\pi k(1+x)}{c^2}, \qquad 17$$

$$\{\frac{\omega^2}{n^2 c^2}[1 - 3\gamma(1+x)] - 1\}^2 = \frac{9\gamma^2\omega^4}{n^4 c^2}(1+x)^2. \qquad 18$$

The dispersion equation (18) the existence of plane waves of two types: $\omega/nc = \pm 1$ and

$$\frac{\omega}{nc} = \pm \frac{1}{\sqrt{1 - 6\gamma(1+x)}}, \qquad (19)$$

where the inequality $6\gamma(1+x) < 1$ must hold for the solutions $\varphi$ and $\varphi_1$ to be stable. Even more stringent condition is required for the wave velocity (19) to be below the speed of light in vacuum:

$$6\gamma(1+x) < 0. \qquad (20)$$

From (16) and (17) at $\lambda_0 = 0$ and when $x \to 0$ for $\varphi_s = \varphi_{1s}$, we obtain Newton's law of gravitation that corresponds to a stationary matter density distribution $\rho_0 = \varepsilon_0 c^{-2}$ for an arbitrary form of the function $\varepsilon_0(\mathbf{r})$.

**3.** Note that the two types of wave solutions obtained above are possible only if $\varphi_1 \neq \varphi$. Indeed, at $\varphi_1 = \varphi$, system (12)-(14) is already closed and there is no need to establish the additional relation between $\varepsilon$ and $p$. In this case, the exact solution of system (12)-(14) is (at $\varphi = \varphi_s + \tilde{\varphi}$)

$$\varepsilon = \varepsilon_0(\mathbf{r}) + \frac{\gamma}{2\pi k} \frac{\partial^2 \tilde{\varphi}}{\partial t^2}, \quad p = \frac{\lambda_0 c^4}{4\pi k} + \varepsilon_0(\mathbf{r}) - \frac{\gamma}{2\pi k} \frac{\partial^2 \tilde{\varphi}}{\partial t^2},$$

$$\Delta \varphi_s = \frac{16\pi k \varepsilon_0(\mathbf{r})}{c^2} + 2\lambda_0 c^2, \quad \Delta \tilde{\varphi} - \frac{1}{c^2} \frac{\partial^2 \tilde{\varphi}}{\partial t^2} = 0, \qquad (21)$$

where the velocity of the gravitational waves $\tilde{\varphi}$ for any $\gamma$ coincides with the speed of light.

**4.** Thus, the interaction between the vacuum and nonvacuum phases of the states of matter at $\gamma \neq 0$ in (14) can manifest itself in the existence of two types of low-amplitude gravitational waves. Their propagation velocity equals to the speed of light for the first type and, in accordance with (19), equals to $v_g = c/n_0$. Here, $n_0 = \sqrt{1 - 6\gamma(1+x)}$ is the effective refractive index of the "medium" different from unity at $\gamma \neq 0$ and $x \neq -1$. According to [36,37], it is at $n_0 \neq 1$ (at both $n_0 > 1$ and $n_0 < 1$), only for gravitational waves of the second type, that an analog of Cherenkov radiation can be realized at above-threshold velocities of ordinary matter when $V > V_{th} = c/n_*$, where

$$n_* = n_0 + \sqrt{n_0^2 - 1}, \; (n_0 > 1); \; n_* = \frac{1 + \sqrt{1 - n_0^2}}{n_0}, \; (n_0 < 1).$$

In this case, the vacuum as a semblance of ether can act as "medium". An experimental study of this emission of gravitational waves of the second type could also make it possible to estimate $\gamma$ when $\varphi \neq \varphi_1$. The exact cosmological solution obtained in the next section (see (42)) corresponds to $\gamma = 1/3$. For $\gamma = 1/3$, condition (20) can be met only at $x < -1$. However, the coincidence of $\gamma$ is by no means necessary for the description of the gravitational waves in (12) and (13) in this cosmological solution as well.

## 4. Accelerated expansion of the Universe without singularities

Based the modification of the general relativity equations (2) and (3) derived from the variational principle (1), in this section we obtain a new scenario for the evolution of the Universe at $\gamma = 1/3$, in which there is no Bid Bang singularity and the mechanism of the observed accelerated cosmological expansion is explained.

**1.** For the modification of the Friedman homogeneous and isotropic model in a comoving frame, from (2) and (3) we obtain

$$\dot{H} = -\frac{4\pi k(\varepsilon + p)}{c^4} + \frac{r}{a^2}, \qquad 22$$

$$\dot{H} = -\frac{4\pi k(p - \varepsilon_d)}{c^4} - \frac{3}{2}H^2 - \frac{r}{2a^2}, \qquad 23$$

$$\varepsilon_d = \frac{c^4}{8\pi k}[\lambda_0 + 3\gamma(2\dot{H} + 3H^2)]. \qquad 24$$

Here $a$ is the scale factor, $H = \dot{a}/a$, parameter $r = 0, \pm 1$ corresponds to zero, positive and negative curvatures of a space, $\varepsilon_d$ in (24) is defined by (5), and we used that $g \propto a^6$. In contrast to the physically justified derivation of system (22)-(24), and outwardly similar model with

$$\varepsilon_d = \frac{3c^4}{8\pi k}(\alpha \dot{H} + \beta H^2),$$

that corresponds to (24) at $\lambda_0 = 0, \alpha = 2\gamma, \beta = 3\gamma$ (i.e., when $\beta = 3\alpha/2$), was formally introduced in [28] at $r = 0$. In this case, the contradiction between the theoretical consideration used (see Eqs. (3) in [28]) and the finiteness of the right-hand side of Eq. (6) with $\varepsilon_d \neq 0$, when $\alpha \neq 0$ and $\beta \neq 0$, is admitted in [28]. The two free parameters, $\alpha$ and $\beta$, are used in [28] for best fitting between the model of dark energy $\varepsilon_d$ and the data set of cosmological observation (SnIa, BAO, and CMB-WMAP). According to [28], these values are

$$\alpha = 0.587 + \frac{0.142}{-0.115}, \quad \beta = 0.949 + \frac{0.151}{-0.110},$$

when the best quantitative agreement (within $1\tilde{\sigma}$, i.e., one root-mean-square deviation) is provided.

The parameter $\gamma = 1/3$, at which $\alpha = 2/3 \approx 0.666$ and $\beta = 1$, falls within this $1\tilde{\sigma}$ range to indicated values. Below (see paragraph 5), we will show that it is at $\gamma = 1/3$ and $r = 0$ that system (22)-(24) admits an exact solution (see below (42)) which agrees well with the data of cosmological observations (see Section 6).

**2.** At arbitrary $\gamma$ and $r$, system (22)-(24) for $\lambda_0 = 0$ and $p = x\varepsilon$ has the invariant

$$I = (\dot{a} + \mu)(\frac{a_0}{a})^{2(1-\theta_0)}, \qquad 25$$

where $a_0 \equiv a(0)$,

$$\mu = \frac{r(1+3x)}{1+3[x-\gamma(1+x)]}, \quad \theta_0 = \frac{3(1+x)(1-3\gamma)}{2[1-3(1+x)]}.$$

An invariant of a similar type was first obtained in [19] for the model (3') with $\lambda_0 = 0$.

In view of invariant (25), under conditions $\theta_0 < 1$ and $\mu > 0$, the following bound on $a$ exist for system (22)-(24) at any $r$ :

$$a > a_{\min} = a_0 \left( \frac{\mu}{\dot{a}^2 + \mu} \right)^{1/2(1-\theta_0)} \qquad 26$$

This means that a singular regime for evolution of system (22)-(24), when $a \to 0$, cannot be realized. At $\gamma \neq 0$ in (24), $a_{\min}$ in (26) is finite only for $r > 0$ and with inequality

$$\frac{1+3x}{3(1+x)} > \gamma > \frac{1}{3(1+x)}, \qquad 27$$

which is possible only at $x > 0$. At the same time, for $\gamma = 0$ the absence of singularity is admissible only at $r > 0$ and for $x < -1/3$. The latter inequality corresponds to violation of the strong energy dominance condition in the well-known theorem on the singularity in general relativity [15]. Thus, we obtained a new proof of this theorem, simpler than that in [15], by using invariant (25).

At any $\gamma \neq 1/3$ and $x = 0$, according to (25) and (26), the singularity turns out to be possible for system (22)-(24). At the same time, for $\gamma = 1/3$ and $x \to 0$, there is no singularity in this system at $r > 0$ and $x > 0$ (or $r < 0$ and $-1/3 < x < 0$) and even at $r = 0$ or $x = -1/3$, when $\mu = 0$ and, in view of (25), the exponential solution for $a(\tau)$ takes place.

**3.** Thus, at any $r$, the parameter $\gamma = 1/3$ turns out to be preferential for system (22)-(24). At $\gamma = 1/3$, this system takes the following relatively simple form:

$$\dot{H} = -\frac{4\pi k(\varepsilon + 3p)}{c^4} + \lambda_0, \qquad 28$$

$$\frac{8\pi k p}{c^4} = \lambda_0 - \frac{r}{a^2}. \qquad 29$$

In particular, at $\varepsilon + 3p = \varepsilon_0 + 3p_0 = const$, system (28), (29), for any $r$, has exact solution:

$$a(\tau) = a_0 \exp\{H_0 \tau + \frac{\tau^2}{2}[\lambda_0 - \frac{4\pi k}{c^4}(\varepsilon_0 + 3p_0)]\}, \qquad 30$$

where $H_0 \equiv H(0) > 0$, because $\tau = 0$ corresponds to the present epoch. With this solution for $a(\tau)$ in (30), $p$ can be determined from (29) and $\varepsilon(\tau) = \varepsilon_0 + 3p_0 - 3p(\tau)$. The scale factor $a(\tau) \to 0$ in (30) with $\tau \to \pm\infty$ if

$$\tilde{\Omega}_{0m} < \frac{2\lambda_0}{3H_0^2} \equiv \Omega_{th}, \qquad 31$$

where

$$\tilde{\Omega}_{0m} = \frac{\tilde{\rho}_0}{\rho_c}, \quad \tilde{\rho}_0 = \frac{\varepsilon_0 + 3p_0}{c^2}, \quad \rho_c = \frac{3H_0^2 c^2}{8\pi k}, \quad \varepsilon_0 = \varepsilon(0), \quad \rho_0 = \rho(0).$$

In contrast, when (31) is violated, there is an asymptotic infinite increase $a \to \infty$ with $\tau \to \pm\infty$. For particular value $\lambda_0 = 3\Omega_{0m}H_0^2$ (where $\Omega_{0m} = \rho_0/\rho_c$, $\rho_0 = (\varepsilon_0 + p_0)/c^2$), we get $\Omega_{th} = 2\Omega_{0m}$ and (31) holds at $\varepsilon_0 > p_0$, which is characteristic of the present epoch.

**4.** According to observational data (see, e.g., reviews [4,5]), the space is flat Euclidian with a high accuracy and we may set $r = 0$ in (22)-(24) and (29). In this case, for any $\gamma$, from (23) we obtain equation

$$(\dot{H} + \frac{3}{2}H^2)(1 - 3\gamma) = \frac{\lambda_0}{2} - \frac{4\pi k p}{c^4}, \qquad 32$$

which, in particular, at $\gamma = 1/3$ leads to (29) for $r = 0$.

Let the right-hand side of (32) be zero for any $\gamma$:

$$\lambda_0 = \frac{8\pi k p}{c^4}. \qquad 33$$

If $\gamma \neq 1/3$, then from (32) and (33) we obtain exact solution

$$a = a_0(1 + \frac{3}{2}H_0\tau)^{2/3}, \qquad 34$$

which does not depend on $\gamma$. According to (22), an evolution of the quantity $\Omega(\tau) = (\varepsilon + p)/c^2\rho_c$ in the form

$$\Omega(\tau) = (1 + \frac{3}{2}\tau H_0)^{-2} \qquad 35$$

corresponds to solution at $r = 0$, for which the conservation law $\tilde{E} = \Omega a^3 = const$ holds, where $\varepsilon$ is a function of $\tau$ and the constant $p$ is determined by (33). Similar conservation law for energy (with $p = 0$, $\beta = 2\gamma$ in (3') and $\lambda_0 = 0$) was obtained in [19].

For solution (34), the accelerated regime of cosmological expansion is absent ( because $\ddot{a} < 0$ ) and a singularity took place for $\Omega(\tau)$ in (35) at $t = t_s = -2/(3H_0c)$.

At $r = 0$ and any $\gamma$, we can obtain a general representation for solution from (22):

$$a(\tau) = a_0 \exp[H_0\tau - \frac{3}{2}H_0^2 \int_0^\tau d\tau_1 \, \tilde{y}(\tau_1)], \qquad 36$$

where $\tilde{y}(\tau) = \int_0^\tau d\tau_1 \Omega(\tau_1)$. A function $p(\tau)$ is determined by (32) with $a(\tau)$ from (36).

In general case, from (32) and (36) we can obtain the following relation between $\Omega_p = p/\rho_c c^2$ and $\Omega$:

$$\Omega_p = \frac{\lambda_0}{8\pi k \rho_c} + (1 - 3\gamma)[\Omega - \frac{3}{2}H_0 \tilde{y}], \qquad 37$$

where $\Omega = \Omega_\varepsilon + \Omega_p$, $\Omega_\varepsilon = \varepsilon/\rho_c c^2$. At $\gamma = 1/3$, Eq. (37) is reduced to (33). Note, only the condition $r = 0$ (i.e., the space is Euclidian) is needed to obtain (37).

It follows from (37) that to use the representation of the equation of states in the form $p = x\varepsilon$ with $x = const$, it is necessary that the function $y(\tau) = \int_0^\tau d\tau_1 \Omega_\varepsilon(\tau_1)$ satisfy the equation

$$x(\dot{y} - \frac{3}{2}H_0 y) \ddot{y} = (\dot{y} - \frac{3}{2}H_0)[2(\dot{y} + \Omega_\lambda) + (x+1)F_0], \qquad 38$$

where

$$\Omega_\lambda = \frac{\lambda_0 c^2}{8\pi k \rho_c}, \quad F_0 = 2(1+x)(1-3\gamma)(\dot{y} - \frac{3}{2}H_0 y)^2 - \dot{y}.$$

In particular, for the equation of state with $x = -1$ to hold, it follows from (38) that

$$(\dot{y} - \frac{3}{2}H_0 y)^2 (\dot{y} + \Omega_\lambda) = C_1, \qquad 39$$

where $C_1$ is an arbitrary constant. For example, at $C_1 = 0$, Eq. (39) is satisfied with $\Omega_\varepsilon(\tau) = -\Omega_\lambda = const$ (at $\lambda_0 < 0$) or with $\Omega_\varepsilon(\tau) = \Omega_\varepsilon(0)\exp(\frac{3}{2}H_0 \tau)$.

Thus, for a flat space (i.e., at $r = 0$), the equation of state with a constant (in time) $x$ can hold only for a certain form of the dependence $\Omega_\varepsilon(\tau)$ that satisfies the nonlinear differential equation (38). In this case, the law of cosmic evolution is defined by (36) for the function $\Omega_\varepsilon(\tau)$ determined from (38).

Irrespective of the existence of relations between $\varepsilon(\tau)$ and $p(\tau)$, solution (36) at $\varepsilon + p > 0$, $H_0 > 0$, $r = 0$ and any $\gamma$ describes the regimes of cosmological expansion ($\dot{a} > 0$ at $\tau < \tau_{max}$) and contraction ($\dot{a} < 0$ at $\tau > \tau_{max}$), with $\tau_{max}$ determined from (36) and the condition $\dot{a}(\tau_{max}) = 0$ using equation

$$H_0 \int_0^{\tau_{max}} d\tau \Omega(\tau) = H_0 \tilde{y}(\tau_{max}) = 1. \qquad 40$$

For the accelerated regime of expansion observed at the present epoch (when $\ddot{a}(0) > 0$ in (36)) to be realizable, the following inequality must hold:

$$\Omega_{0m} = \frac{\rho_0}{\rho_c} < \frac{2}{3}, \qquad 41$$

where $\Omega_{0m} = \Omega(0)$.

According to observational data, $\Omega_p \to 0$, $\tilde{\Omega}_{0m}$ actually coincide with $\Omega_{0m}$ and equal to $\Omega_\varepsilon(0) < 0.3$, which satisfies (41) at any $\gamma$ and irrespective of the specific form of the functions $\varepsilon(\tau)$ and $p(\tau)$ at $\tau \neq 0$.

**5.** Let us now consider separately the special case with $\gamma = 1/3$, $r = 0$ in system (22)-(24). In this case, it follows from (23) and (32) that $p = p_0 = const$ and correspond to (33) or (37) (at $\gamma = 1/3$ in (37)). The general solution of system (22)-(24) is (36) with $p$ from (33) and for any function $\varepsilon(\tau)$. Note that, in this case, Eq. (6) is satisfied identically, retaining the arbitrariness of the function $\varepsilon(\tau)$. In particular, solution (36)

with $p$ from (33) and $\varepsilon(\tau) = \varepsilon_0 = const$ is admissible. In this case, (36) becomes

$$a(\tau) = a_0 \exp(H_0\tau - \frac{2\pi k \rho_0}{c^2}\tau^2), \qquad 42$$

where $\rho_0 = (\varepsilon_0 + p_0)/c^2 = \Omega_{0m}\rho_c$ and $p_0 = \lambda_0 c^4/8\pi k$, according to (33). Solution (42) coincides with (30), in view of (33). In contrast to (36), where $p = p_0$ and $\varepsilon(\tau)$ is an arbitrary function, solution (42) seems more natural for arbitrary relations between constant (in time) quantities $\varepsilon_0$ and $p_0$.

For solution (42), $\tau_{max} = 2/3H_0\Omega_{0m}$ in Eq. (40). According to (42), the cosmological expansion at $t < t_{max} = \tau_{max}/c$ and contraction at $t > t_{max}$ occurs with a positive acceleration ($\ddot{a}(\tau) > 0$) only outside the interval

$$t_{max}(1 - \varpi) \leq t \leq t_{max}(1 + \varpi), \quad \varpi = \sqrt{\frac{3}{2}\Omega_{0m}}. \qquad 43$$

Inside this interval, on the contrary, $\ddot{a} < 0$. For example, at $\rho_0 \approx 0.23 \times 10^{-29} g\ cm^{-3}$ and $H_0 c \approx 0.24 \times 10^{-17}\ s^{-1}$ [4], we obtain an estimate $t_{max} \approx 39.6 \times 10^9 yr$. In this case, according to (43), it remains $t_{max} - t_0 \approx 16.7 \times 10^9 yr$ (where $t_0 = 1/H_0 c\varpi \approx 22.9 \times 10^9 yr$) until the end of the period of accelerated expansion from the present epoch (at $t = 0$).

The exact solution (42) corresponds to the possibility of an unlimited existence in time (from $t \to -\infty$ to $t \to \infty$) for a Universe without any singularities in the past and the future. In the Appendix, we show, however, that solution (42) is stable only in the regime of expansion at $t < t_{max}$, while the corresponding regime of contraction at $t > t_{max}$ is exponentially unstable against small metric perturbations. This can imply the need for considering the change of regime (42) at $t > t_{max}$ to a different evolutionary regime with a different value of the parameter $\gamma$ or with the more general model (3') from [19].

Note that a decrease in the acceleration of the expansion of the Universe with $\ddot{a}(0) < 0$, which, according to (42), gives inequality

$$\Omega_{0m} > 2/8 \approx 0.22\ldots, \qquad 44$$

corresponds to the observational data. Condition (44) is satisfied for the observed quantity $\Omega_{0m} \leq 0.3$, which falls into the interval of values established by inequalities (41) and (44).

For comparison with observational data (see Section 6), it is convenient to use the redshift $z = a_0/a(\tau) - 1$ instead of $a(\tau)$. In this case, the following form of the function

$$h(z) = H(z)/H_0 = \sqrt{1 + 3\Omega_{0m}\ln(1 + z)} \qquad 45$$

corresponds to solution (42). In particular, we can determined $\dddot{a}$ for any $z$, by using (45), in the form:

$$\frac{\dddot{a}}{a_0 H_0^3} = \frac{h(z)}{1+z}\left\{1 + 3\Omega_{0m}[\ln(1+z) - \frac{3}{2}]\right\}. \qquad 46$$

It follows from (46), for example, at $\Omega_{0m} = 0.3$ that the observed (see [6]) slowdown of the acceleration of cosmological expansion should take place only at $z < 0.475$, while $\ddot{a} > 0$ should be for $z > 0.475$.

## 5. The scalar-field model of dark energy with a negative cosmological constant

Let us show that the exact solution (42) obtained above for a spatially flat model (with $r = 0$ in system (22)-(24)) can coincide with the exact solution of a new generalization of the scalar-field model [26] under certain condition. In particular, it follows from the condition of such coincidence that the mass of the ultralight particles produced out of vacuum during cosmological expansion can be estimated.

**1.** Consider the simplest equation to describe the scalar field $\Phi$ to which particles with mass $m$ correspond [2,26]:

$$\ddot{\Phi} + 3H\dot{\Phi} + \overline{m}^2 \Phi = 0, \quad \overline{m} = mc\hbar^{-1} \qquad 47$$

We will use Eq. (47) in combination with system (22), (23) and formal replacement of $\varepsilon$ and $p$ in this system (at $r = 0$) by

$$\varepsilon_\Phi = \frac{1}{2}(\dot{\Phi}^2 + \overline{m}^2 \Phi^2) \quad \text{and} \quad p_\Phi = \frac{1}{2}(\dot{\Phi}^2 - \overline{m}^2 \Phi^2),$$

(see, e.g., [38]). We will also use the constant quantity $\tilde{\varepsilon}_{do} = \tilde{\lambda}_0 c^4/8\pi k$ instead of $\varepsilon_d$. As a result, we obtain a scalar-field model in which the field $\Phi$ can, as usual [2], describe dark energy. At the same time, as we show below, the cosmological constant $\tilde{\lambda}_0$ at its negative value corresponds to a constant density of dark matter being continuously produced out of vacuum during cosmological expansion. Note that system (22), (23), (47) at $\tilde{\lambda}_0 = 0$ and $r = 1$ coincides with that considered in [26], where an approximate, at $mc\tau/\hbar \gg 1$ (see (10) in [26]), asymptotic solution with $\Phi = B\tau$ at $B = const$ was obtained. In contrast to [26], we obtain the following exact solution at $r = 0$ and $\tilde{\lambda}_0 < 0$ for (22), (23), (47):

$$\Phi = B\tau + B_0, \quad B_0 = -\frac{H_0 c^4}{8\pi kB}, \quad \tilde{\lambda}_0 = -\frac{4\pi kB^2}{c^4}, \qquad 48$$

$$\overline{m}^2 = \frac{12\pi kB^2}{c^4}, \qquad 49$$

$$a(\tau) = a_0 \exp(H_0 \tau - \frac{2\pi kB^2 \tau^2}{c^4}). \qquad 50$$

**2.** Solution (50) for $a(\tau)$ can coincide with the exact solution (42) obtained above if the following equality holds:

$$\rho_0 c^2 = B^2, \qquad 51$$

where $\rho_0 = (\varepsilon_0 + p_0)/c^2$ at $p_0 = \lambda_0 c^4/8\pi k$, according to (33). Generally, $\varepsilon_0$ and $p_0$ in (42) are not related in any way if $\lambda_0$ is not related to $\tilde{\lambda}_0$ from (48). However, when the relation $\tilde{\lambda}_0 = \alpha\lambda_0$ holds, where $\alpha$ is an arbitrary dimensionless constant, we obtain the following equation of state under condition (51):

$$p_0 = x_0\varepsilon_0, \quad x_0 = -1/(1+2\alpha). \qquad 52$$

At the exact coincidence $\lambda_0 = \tilde{\lambda}_0$, i.e., at $\alpha = 1$, we obtain the equation of state with $x_0 = -1/3$ from (52) that can correspond to a nonequilibrium gas of scalar photon pairs being produced out of the vacuum (see subparagraph 4). According to (52), the ordinary ultrarelativistic equation of state with $x_0 = 1/3$ is obtained at $\alpha = -2$ [32].

Thus, condition (510) is necessary for the exact solutions (50) and (42) to coincide both in the presence and in the absence of a relation between $\lambda_0$ and $\tilde{\lambda}_0$ as well as between $p_0$ and $\varepsilon_0$ in (52). Condition (51) leads to the fulfillment of the equalities $\varepsilon + \varepsilon_d = \varepsilon_\Phi + \tilde{\varepsilon}_d$ and $p - \varepsilon_d = p_\Phi - \tilde{\varepsilon}_d$, which are necessary for the coincidence of the Einstein forms of the equations [3] considered for various cosmological models with differing physical grounds.

**3.** The quantity $B$ in solution (48)-(50) is related to the particle mass by (49). Therefore, the following relation between the constant density $\rho_0$ of dark matter and the mass $m$ of the particles being produced out of the vacuum during cosmological expansion can be obtain from condition (51):

$$m = m_0 = \frac{\hbar}{c^2}\sqrt{12\pi k\rho_0}. \qquad 53$$

Here $m_0 = \hbar\sqrt{3}/c^2 t_0 \approx 3 \times 10^{-66}$g at the values of the parameters used above in connection with inequality (43). The quantities $\rho_0$ and $\lambda_0$ turn out to be interrelated when (52) holds: $\rho_0 = q\lambda_0 c^2/4\pi k$.

The derived relation (53) between $m_0$ and $\rho_0$ differs qualitatively from that in [10], where $\varepsilon_d = km^2/\lambda_m^4$ (with $\lambda_m = \hbar/mc$) is estimated by taking into account the contribution from the gravitational energy of particles to the vacuum energy density $\varepsilon_d = \lambda_0 c^4/8\pi k$. In [12], the role of such gravitationally active particles is fulfilled by vacuum virtual photons.

Here, in contrast to [10,12,31], the realizability of the self-generation of particles with mass $m$ out of the vacuum through the interaction between the vacuum and nonvacuum phases of the state of matter, according to (42), leads to condition (53).

**4.** Representation (53) can correspond to a finite rest mass of real photons. Indeed, based on the uncertainty principle, the authors of [39] obtained an estimate for the photon rest mass that is quantitatively close to (53):

$$m = \frac{\hbar}{\Delta t c^2} = \frac{3.7 \times 10^{-66}g}{T},$$

where $T$ is the lifetime of the Universe in units of $10^{10}yr$. In addition, the estimate of $m_0$ in (53) is close in order of magnitude to the mass of the hypothetical inflaton [2] (when $mc/\hbar = 3H_0/2$ in (47)) and is consistent with the existing experimental estimates of the upper limit for the photon rest mass[2)] [39-43]. On the other hand, Eq. (47) corresponds to the description of a scalar particle and can not be used to directly describe a single photon. However, it can be used to characterize a scalar photon pair that consists of two photons with a total zero helicity and flying apart at a finite angle $\theta \neq 0$ with respect to each other. According to [41], the rest mass of such a system is $\tilde{m}_0 = (2\hbar v/c^2)\sin\theta$, where $v$ is the frequency of each of these photons that constitute the scalar photon pair. For example, at an appropriate CMB frequency, $v \approx 0.5 \times 10^{10} s^{-1}$, we obtain $\theta \approx 10^{-28} rad$ from the condition $\tilde{m}_0 = m_0$ (for $m_0$ from (53)). As a result, to separate by 1 cm from each other, these photons should overcome a distance of $10^{28} cm$, comparable to the size of the visible Universe. However, this hypothesis does not affect in any way the entire analysis performed here and does not rule out the possibility of the existence of other dark matter sources. This can be not only the individual finite rest mass of photons but also the effective rest mass $m_{eff}$ for a photon in a medium [37,38,42], when, for example, $m_{eff} \approx 10^{-50} g$ for the intergalactic medium [42].

Above, we pointed out the important role of the finiteness of not only $\dot{\varepsilon}_d \neq 0$ at $\gamma \neq 0$ but also the cosmological constant $\lambda_0$ at both cases $\lambda_0 > 0$ and $\lambda_0 < 0$. The physical meaning of these inequalities can be additionally refined if we use the following relation between the rate of change in the density of distributed source-sinks $\dot\sigma$ and $\lambda_0$ for solution (42):

$$\dot\sigma = -\frac{3(1+x_0)}{2x_0}\lambda_0,$$

where we took into account that $\dot\sigma = 3\ H$ for solution (42). In particular, for the case of $x_0 = -1/3$ noted above (at $\alpha = 1$ in (52)), we have

$$\lambda_0 = \frac{\dot\sigma}{3} = -\frac{4\pi k\rho_0}{c^2} < 0,$$

when $\lambda_0 = \dot H_0$. For comparison, note that the relation $\lambda_0 = 3H^2$ holds for the de Sitter solution [25].

The equation of state at $x_0 = -1/3$ (which also emerges in the string theory [2,44]) can correspond to the nonequilibrium process of the formation of a new phase from the vacuum phase of matter [9], when something (say, massive scalar photons pairs) can be produced out of the vacuum, just like cavitation air bubbles.

## 6. Comparison with observational data

Let us consider the representation of the function $h(z)$ in (45) for the exact solution (42) and use it for comparison with the data of present-day cosmological observation that are characterized by the defined by the form of $h(z)$ [3].

**1.** In accordance with (45), the quantity $\ddot a$ in (46) is defined for any $z$. An evolutionary cosmological regime with a slowdown of the accelerated expansion of the Universe, $\ddot{a} < 0$, was shown to be possible for $z < 0.475$ at $\Omega_{0m} \approx 0.3$. This is consistent with the available observational data [6].

**2.** Based on (45), we can estimate quantity

$$A = \frac{\sqrt{\Omega_{0m}}}{h^{1/3}(z_1)} \left[ \frac{1}{z_1} \int_0^{z_1} \frac{dz}{h(z)} \right]^{2/3}$$

at $z_1 = 0.35$ that corresponds to the baryonic acoustic peak measurements (SDSS BAO data), $A = 0.469 \pm 0.017$ [3,45]. For $h(z)$ from (45) and $\Omega_{0m} = 0.28$, we obtain a close value $A \approx 0.49$. For another quantity $d = D_v(0.35)/D_v(0.20)$, where $D_v(z) = zA/H_0\sqrt{\Omega_{0m}}$, according to the observations [45], we have $d = 1.812 \pm 0.060$. For the exact solution (45) without any fitting parameter (except the same $\Omega_{0m} = 0.28$), we obtain a close value of $d \approx 1.7$.

**3.** The following function (see (2.1) in [7]) is used to analyze the observations of supernovae SnIa:

$$\mu(z) = m - M = 5\log_{10}\left(\frac{d_L}{M_{pc}}\right) + 25, \quad d_L = \frac{1+z}{H_0} \int_0^z \frac{dz'}{h(z')}.$$

Here $\mu$ is the distance module, $m$ and $M$ are apparent and absolute magnitudes of the source correspondingly and $d_L$ is the luminosity distance. The figure compares the dependence $\mu(z)$ that we derived at $\Omega_{0m} = 0.3$ for $h(z)$ from (45) with the observational data used in [7] and with $\mu(z)$ for other theoretical models [7]. We see from the figure that the exact solution (45) for the presented $z$ range agrees well with the observational data.

Thus, for relatively low $z$, the exact solution (42) in representation (45) provides conclusions that agree well with the data of model-independent observations, i.e., with the data that are not related to any model theoretical conclusions (as, e.g., for the data at $z > 10^3$, when the principles of the theory of a hot Universe are used explicitly). Note that in [6] attention is also drawn to some dissonance of the data at $z < 1$ and $z > 10^3$.

## 7. Conclusions

Here, we obtained a new modification of the general relativity equations that corresponds to a generalization of the various aspects of the theory of vacuum in general relativity [8,9,11]. The exact solution (42) obtained on its basis admits unlimited (see also [46]) existence in time of a spatially flat Universe without any singularities that correspond, for example, to the Big Bang. During cosmological expansion with an acceleration changing with time (see (46)), solution (42) is stable (see the Appendix). A constant density $\rho_0$ of ordinary (dark) matter due to continuous production of ultralight (see (53)) particles of this matter (possibly, in the form of scalar massive photon pairs) out of the vacuum corresponds to it. In this case, the problem of cosmological constant [1,4] is naturally removed. The presence of a new fundamental mass $m_0$ connected to the invariant $\rho_0$ (see (53)) admits a relation $\rho_0 = \rho_P(m_0/m_P)^2$, where $\rho_P$ is the Plank density and $m_P \approx 10^{-6}g$. This is distinct from the relation considered in [4,47], $\rho_0 = \rho_P(M_{EW}/m_P)^8$, where $M_{EW}$ is the mass that corresponds to electroweak interaction. The quantity $m_0$ defined in (53) can also serve as a basis for subsequently reconsidering the problem of divergence in quantum electrodynamics [48,49], irrespective of a particular interpretation of the quantity $m_0$ itself.

### Remarks

**1)** Matter composed of particles with a nonzero rest mass for which the comoving frame of reference can be determined uniquely is considered to be ordinary [8,9]. For matter in the vacuum phase any frame of reference is comoving [8].
**2)** The influence of photon oscillation on the slope of the cosmic microwave background (CMB) spectrum was considered in [42], while the absence of mixed polarization states for the circularly polarized waves that correspond to longitudinal massive photons is pointed out in [40], which is important in analyzing the CMB polarization observations.

### Acknowledgments

We are grateful to V. N. Lukash, V. N. Strokov, and E. N. Pogorelov for helpful discussions and useful remarks and to A. G. Chefranov for help in preparing the materials used to compare the theory with observational data. We thank the referees (JETP and Letters to JETP, where an abbreviated version of this paper was considered and our attention was drawn to the paper [26]) for benevolence and constructive remarks.

### Appendix: Stability of the solution

To analyze the stability of the dynamical regime that correspond to the exact solution (42) (of system (22)-(24) at $r = 0$, $\gamma = 1/3$), we obtain the following equations for small perturbations in metric ($g_{ik} \to g_{ik} + h_{ik}$, $h = h_i^i$) from (2) and (3):

$$\ddot{h}^{\beta}_{\alpha} + 3H\,\dot{h}^{\beta}_{\alpha} + \frac{1}{a^2}(h^{\gamma;\beta}_{\alpha;\gamma} + h^{\beta;\gamma}_{\gamma;\alpha} - h^{\beta;\gamma}_{\alpha;\gamma} - h^{;\beta}_{;\alpha}) = 0, \quad \alpha \ne \beta, \qquad \text{A.1}$$

$$\ddot{h}\,[1 - 3\gamma(1+p_\varepsilon)] + 3H(1+p_\varepsilon)(1-3\gamma) + \frac{h^{\gamma;\alpha}_{\alpha;\gamma} - g^{;\alpha}_{;\alpha}}{2a^2}(1+3p_\varepsilon)$$

$$- \frac{3\gamma c^4 (1+p_\varepsilon)}{8\pi k a^2 (p+\varepsilon)}\left[ 3H(\dot{h}^{;\alpha}_{;\alpha} - \dot{h}^{\alpha;\beta}_{\beta;\alpha}) + \ddot{h}^{;\alpha}_{;\alpha} - \ddot{h}^{\alpha;\beta}_{\beta;\alpha}\right], \qquad \text{A.2}$$

where $H = \dot{a}/a$, $p_\varepsilon \equiv \partial p/\partial\varepsilon$. Eqs. (A.1) and (A.2) at $\gamma = 0$ reduced to the system (115.6) from [32] (here, in distinction to [32], where the derivatives with respect to $\eta$, $(d\tau = ad\eta)$ are used, we use differentiation with respect to $\tau$, which is denoted by the dot, as in the main text).

From (A.1) and (A.2) for perturbations in the form of plane waves,

$$h^{\beta}_{\alpha} = \left[\lambda\left(\frac{\delta^{\beta}_{\alpha}}{3} - \frac{n_\alpha n^\beta}{n^2}\right) + \frac{\mu}{3}\delta^{\beta}_{\alpha}\right]\exp(i\mathbf{nr}), \quad h = \mu,$$

we obtain a system for the unknown functions $\lambda$ and $\mu$:

$$\ddot{\lambda} + 3H\,\dot{\lambda} - (\lambda + \mu)\frac{n^2}{3a^2} = 0, \qquad \text{A.3}$$

$$\ddot{\mu}\left[1 - 3\gamma(1+p_\varepsilon) + \tilde{\gamma}\right] + \dot{\lambda}\tilde{\gamma} + \dot{\mu}\,H\left[3(1-3\gamma)(1+p_\varepsilon) + 3\,\tilde{\gamma}\right]$$

$$+ 3\,\dot{\lambda}\,H\,\tilde{\gamma} + (\lambda + \mu)\frac{n^2(1+3p_\varepsilon)}{3a^2}, \qquad \text{A.4}$$

where $\tilde{\gamma} = \frac{n^2 c^4 \gamma(1+p_\varepsilon)}{4\pi k a^2 (p+\varepsilon)}$. System (A.3), (A.4) at $\gamma = 0$ (and, consequently, $\tilde{\gamma} = 0$) corresponds to (115.15) in [32]. In the case of the exact solution (42) for $a(\tau)$ at $\gamma = 1/3$, $p + \varepsilon = p_0 + \varepsilon_0 = \rho_0 c^2$ and $p_\varepsilon = -1/3$ (i.e., at $\alpha = 1$ in (52)), we obtain a relatively simple system from (A.3) and (A.4):

$$\ddot{\lambda} + 3\left(H_0 - \frac{4\pi k\rho_0 \tau}{c^2}\right)\dot{\lambda} - \frac{n^2}{3a^2}q = 0, \qquad \text{A.5}$$

$$\ddot{q} + 3\left(H_0 - \frac{4\pi k\rho_0 \tau}{c^2}\right)\dot{q} = -\ddot{\mu}\,\frac{6\pi k\rho_0 a^2}{n^2 c^2}, \qquad \text{A.6}$$

where $q = \lambda + \mu$ and the function $a$ coincides with (42).

In the limit $n \to \infty$ (corresponding to the analysis in [32], when $h^{\beta}_{\alpha}$ and $h$ are sought in the form of plane waves), the term on the right-hand side of Eq. (A.6) can be neglected. In this case, the function $q(\tau)$ decreases exponentially only for the evolution time that corresponds to the regime of cosmological expansion in (42), when $t < t_{\max} = 2/(3H_0\Omega_{0m}c)$. The same conclusion regarding the evolution of the function $\lambda(\tau)$ can be drawn from (A.5), taking into account that the last term in the left-hand side of (A.5) becomes small (at $q \to 0$ or, formally, at $n \to 0$).

At $t > t_{\max}$, there is an exponential instability against small metric perturbations in the limiting regimes (in $n$).

**Figure capture: Comparison of the experimental data with the results of theoretical models.**

The thick solid line represents the exact solution (45) obtained here at $\Omega_M \equiv \Omega_{om} = 0.3$ (the ratio of the entire ordinary matter, including the dark one, to the critical density). The thin solid, dotted and dashed lines correspond to the three models with different $\Omega_M$ and $\Omega_\Lambda \equiv \Omega_\lambda$ (the ratio of the dark energy density to the critical density). The circles indicate the observational data of two teams of researches.

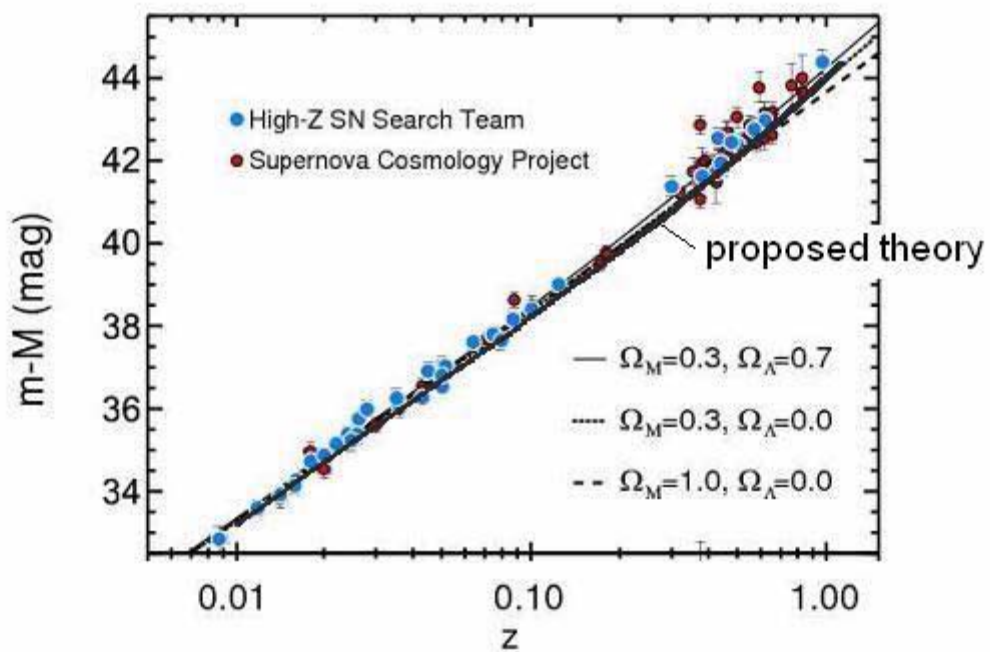